\documentclass[acmtog]{acmart}
\newcommand{\status}{3}

\input{conf/packages}
\definecolor{INCOMPLETECOLOR}{RGB}{178,34,34}
\definecolor{UNDERREVISIONCOLOR}{RGB}{210,121,121}
\definecolor{FEEDBACKNEEDEDCOLOR}{RGB}{230,170,50}
\definecolor{FEEDBACKGIVENCOLOR}{RGB}{121,210,121}
\definecolor{COMPLETECOLOR}{RGB}{121,124,210}
\definecolor{LOCKEDCOLOR}{RGB}{153,102,255}

\definecolor{TODOCOLOR}{RGB}{255,0,0}
\definecolor{HENRYCOLOR}{RGB}{0,0,255}
\definecolor{QICOLOR}{RGB}{118,185,0}
\definecolor{COLINCOLOR}{RGB}{127,127,0}
\definecolor{JENNACOLOR}{RGB}{127,0,127}
\definecolor{KENNYCOLOR}{RGB}{127,100,127}
\definecolor{NIALLCOLOR}{RGB}{127,50,0}
\definecolor{PRATHAMCOLOR}{RGB}{0,100,127}

\definecolor{GUESTCOLOR}{RGB}{0,127,127}

\definecolor{WHITE}{RGB}{255,255,255}

\newcommand{\nothing}[1]{}

\newcommand{\Caption}[2]{\caption[#1]{{\em #1} #2}}

\ifthenelse{\equal{\status}{0}} {          
    \newcommand{\revise}[2]{\textcolor{INCOMPLETECOLOR}{\st{#1}}\textcolor{blue}{#2}}
}{}
\ifthenelse{\equal{\status}{1}} {          
    \newcommand{\revise}[2]{#2}
}{}
\ifthenelse{\equal{\status}{2}} {          
    \newcommand{\revise}[2]{\textcolor{blue}{#2}}
}{}
\ifthenelse{\equal{\status}{3}} {          
    \newcommand{\revise}[2]{#2}
}{}
\ifthenelse{\equal{\status}{4}} {          
    \newcommand{\revise}[2]{#2}
}{}

\ifthenelse{\equal{\status}{0}} {
    \newcommand{\badge}[2]{\colorbox{#1}{\small\textcolor{WHITE}{\texttt{#2}}}}
    \newcommand{\headerBadge}[2]{\hspace*{\fill}\badge{#1}{#2}}

    \newcommand{\feedbackNeeded}{\headerBadge{FEEDBACKNEEDEDCOLOR}{feedback needed}}
    
    \newcommand{\incomplete}{\headerBadge{INCOMPLETECOLOR}{incomplete}}
}{
    \newcommand{\badge}[2]{}{}
    \newcommand{\headerBadge}[2]{}{}

    \newcommand{\feedbackNeeded}{}
    
    \newcommand{\incomplete}{}
}

\newcommand{\ecc}{\theta}
\newcommand{\blur}{\sigma}
\newcommand{\sharpen}{\phi}
\newcommand{\fquadratic}{\mathcal{M}_\text{quadratic}}
\newcommand{\flinear}{\mathcal{M}_\text{linear}}
\newcommand{\param}{k}
\newcommand{\ksize}{s}

\begin{document}
\copyrightyear{2026}
\acmYear{2026}
\acmSubmissionID{613}
\citestyle{acmauthoryear}

\setcopyright{cc}
\setcctype{by}

\acmConference[SIGGRAPH Conference Papers '26]{Special Interest Group on Computer Graphics and Interactive Techniques Conference Conference Papers}{July 19--23, 2026}{Los Angeles, CA, USA}
\acmBooktitle{Special Interest Group on Computer Graphics and Interactive Techniques Conference Conference Papers (SIGGRAPH Conference Papers '26), July 19--23, 2026, Los Angeles, CA, USA}
\acmDOI{10.1145/3799902.3811108}
\acmISBN{979-8-4007-2554-8/2026/07}
\title{Dichoptic Foveation}

\author{Henry Kam}
\email{henrykam@nyu.edu}
\orcid{0009-0002-5913-4647}
\affiliation{%
  \institution{New York University}
  \city{New York}
  \country{USA}}
  
\author{Colin Groth}
\email{c.groth@nyu.edu}
\orcid{0000-0001-6445-5563}
\affiliation{%
  \institution{New York University}
  \city{New York}
  \country{USA}}
  
\author{Jenna Kang}
\email{jennakang@nyu.edu}
\orcid{0009-0006-0544-5182}
\affiliation{%
  \institution{New York University}
  \city{New York}
  \country{USA}}

\author{Pratham Saraf}
\email{ps5218@nyu.edu}
\orcid{0009-0001-9901-0625}
\affiliation{%
  \institution{New York University}
  \city{New York}
  \country{USA}}

\author{Qi Sun}
\email{qisun@nyu.edu}
\orcid{0000-0002-3094-5844}
\affiliation{%
  \institution{New York University}
  \city{New York}
  \country{USA}}

\author{Kenneth Chen}
\email{kennychen@nyu.edu}
\orcid{0000-0002-8095-4407}
\affiliation{%
  \institution{New York University}
  \city{New York}
  \country{USA}}
\renewcommand{\shortauthors}{Kam et al.}


\begin{teaserfigure}
    \includegraphics{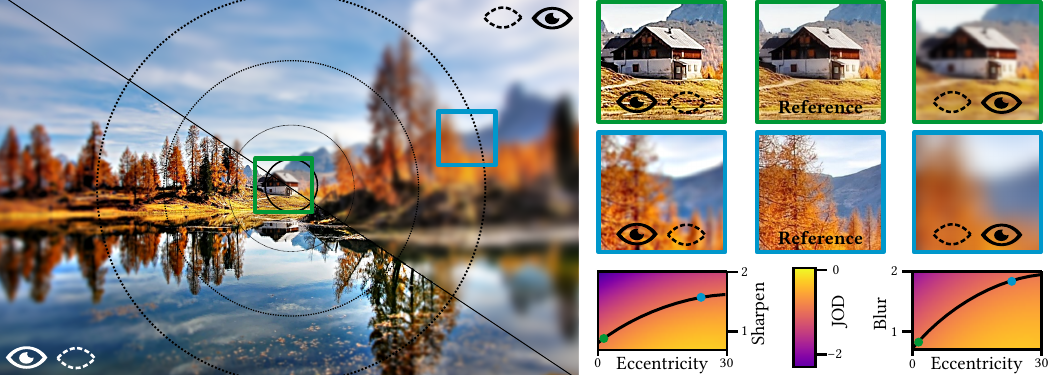}
    \caption{We present dichoptic foveation, a new paradigm to parameterize quality degradation of dichoptically-rendered stimuli in stereoscopic displays like virtual reality. Particularly, we apply Gaussian blur on the stimulus of one eye (right inset images here), while the other eye's stimulus is processed by a simultaneous sharpening routine (left insets).
    This is guided by our computational model (see heatmaps), which is built on psychophysical data. Dichoptic foveation has direct implications for better quality-to-compute ratios in foveated rendering. Image by kordi\_vahle from Pixabay.}
    \label{fig:teaser}
\end{teaserfigure}

\begin{abstract}
Interocular differences in visual perception can induce a variety of effects when fused by the brain.
For example, prior works have found that carefully crafted binocular differences in local detail can improve contrast.
It has also been found that when the frequency content of two stimuli are slightly different, blur suppression leads to a fused percept that is typically dominated by the sharper image.
In this paper, we develop a psychophysical framework to measure the perception of natural image stimuli with interocular frequency differences across the visual field.
To this end, we study the effect of \textit{dichoptic foveation}, which we define as the application of blur to one eye and a simultaneous sharpening filter to the other.
Stimuli were viewed in a virtual reality (VR) head-mounted display (HMD) and placed at different retinal eccentricities.
Study data were scaled to a perceptual just objectionable difference (JOD) scale, and a 4D model was fit to it; our results suggest that interocular frequency differences can be well described by a simple computational model.
We applied the model in a realistic VR scenario with free exploration of 360° videos to improve a base foveated rendering system by enhancing high frequency information dichoptically.
\end{abstract}

\maketitle


\section{Introduction \incomplete}
\label{sec:introduction}


Understanding how we integrate binocular information is a long-standing problem with direct implications for designing stereoscopic displays like VR HMDs.
For example, this principle has been shown to boost contrast \cite{zhong2019dice}.
A large body of visual psychophysics work has also shown that when the two eyes receive information with unequal spatial fidelity, the fused percept is often dominated by the sharper stimulus.
However, no previous work has explored this fusion effect for stimuli located across the visual field.


Significant blur in one eye can remain undetected when perceived stereoscopically, as shown in studies of interocular blur suppression \cite{georgeson2014binocular,collins1994interocular,schor1987ocular} or binocular contrast differences \cite{wang2023effect}.
This finding suggests that the visual system may preserve high-frequency information even when interocular frequency differences occur.
Prior work on dichoptic rendering has shown that carefully crafting differences in binocular imagery can improve depth perception \cite{wang2021reexamination} or contrast perception \cite{zhong2019dice}. 
However, sensitivity to these interocular differences across retinal eccentricities has not been studied, especially regarding how blur and sharpening across the two eyes interact dichoptically.


In this paper, we study \textit{dichoptic foveation}, which we define as the deliberate presentation of a blurred stimulus to one eye and the sharpened image to the other, and its variation with retinal eccentricity\revise{}{, defined as angular distance from the fovea}, as exemplified in \Cref{fig:teaser}.
We ask the question: at a given eccentricity, can fusing blurred and sharpened images across the two eyes yield equivalent, or even improved perceived quality compared to symmetric presentation?


Based on our psychophysical data, we fit a simple perceptual model that predicts a quality correlate given blur--sharpen parameters and eccentricity.
We then demonstrate how this model can be applied to improve foveated rendering, and evaluate this system via a subjective evaluation study.
Using blur magnitude as a proxy to calculate system-agnostic computational cost, we show that our application of dichoptic foveation in VR can reduce sampling rates by about half that of the baseline foveated rendering technique.
For this comparison, both foveated rendering and dichoptic foveation were calibrated to have minimal perceptual impact compared to the reference full-resolution content. Please refer to our supplementary webpage for all data, video examples, and interactive comparisons.
%
%
To summarize, we
\begin{itemize}[leftmargin=2em]
    \item develop a psychophysical framework and collected a large-scale dataset measuring perception of dichoptic stimuli with 12,989 human subject study trials;
    \item \revise{create a perceptual 4-dimensional model that predicts just-objectionable-difference (JOD) for a given combination of stereoscopic blur and sharpen, as well as eccentricity;}{fit a model over three variables (blur, sharpen, and eccentricity) that predicts quality in just-objectionable-differences (JODs);}
    \item leverage the model as an application to improve foveated rendering, which demonstrates the potential for \textasciitilde50\% compute savings at similar perceptual quality.
\end{itemize}

\section{Background \& Related Work}
\label{sec:related}

First, we describe the scientific background of binocular perception, then we address prior literature on dichoptic and foveated rendering. 

\subsection{Binocular and Dichoptic Perception}

Stereoscopic perception, or stereopsis, emerges when the human visual system (HVS) integrates retinal images from the left and right eyes into a unified percept \cite{cang2023neural,blake2011binocular}. 
The neuropsychological mechanism that enables the combination of two monocular images into a single percept is referred to as binocular fusion, and occurs when light stimulates corresponding regions on the two retinas \cite{sperling1970binocular}.

Fusion occurs in the forward-facing binocular visual field, the region where fields of view (FOVs) overlap between the eyes, which spans approximately $110^\circ$ in the horizontal direction \cite{stidwill2017normal}. Outside of this region, each eye has a $40^\circ$ monocular periphery where no overlap occurs between the two eyes \cite{stidwill2017normal}. It is critical to consider this characteristic of the HVS in dichoptic perception because the input of both eyes is required to form a fused stimulus.

Within the binocular visual field, the HVS can tolerate small interocular differences in sharpness without disrupting fusion. As these differences increase, blur suppression may occur, in which the sharper retinal image dominates \cite{georgeson2014binocular}. 
When interocular discrepancies become sufficiently large, binocular rivalry may occur. Binocular rivalry is often characterized by alternating perceptual dominance between the two monocular images \cite{arnold2007staying}. \citet{riesen2019humans} found that rivalry and fusion are not necessarily mutually exclusive states; a "tristable" zone spans $25^\circ$--$35^\circ$ where left or right eye-dominance could occur.

Dichoptic perception refers to the perception of differing visual inputs between the two eyes. The described processes define the possible regimes of dichoptic perception \cite{jiang2023integration, qian2019dichoptic}.
When dichoptic inputs are sufficiently similar, binocular fusion occurs and yields stereopsis \cite{jiang2023integration}; differences primarily in sharpness with similar structure can result in interocular blur suppression, producing a fused percept dominated by one eye \cite{lew2021stimulus}. Applying blur dichoptically to high-spatial-frequency content has a stronger likelihood of fusion breaks than luminance-contrast changes, especially under divergence stress \cite{dostalek2019influence}.
When dichoptic differences become too large, fusion fails and binocular rivalry emerges, which may lead to alternating monocular dominance \cite{law2013dichoptic}.

\subsection{Dichoptic Rendering in Computer Graphics}

An example application of dichoptic perception to computer graphics is the work of \citet{zhong2019dice}, which presents images tone-mapped differently for each eye. The authors found that the fused image has improved perceived contrast.
Similar results were observed for unintended dichoptic contrast differences in AR \cite{wang2023effect}. 
\citet{hoffman2010focus} investigated how focus cues, like blur and depth-dependent information, are used by the HVS to solve the binocular correspondence problem. 
For guidance tasks in VR, dichoptic stimuli are often used to attract gaze~\cite{kudnick2026rippleVision, kudnick2025warpedVision}.
Recent related work has found that applying different compression rates \cite{lee2025assymetric,fezza2017perceptually,shao2010stereoscopic} to each eye can reduce computation at similar perceptual quality.
No prior work, however, has studied the perception of dichoptic frequency differences across the visual field.

\subsection{Foveated Rendering Techniques}
Human visual acuity decreases with increasing retinal eccentricity due to the variation in density of rods, cones, and retinal ganglion cells across the visual field.
Foveated rendering is a set of techniques taking advantage of this relationship by reducing rendering samples with eccentricity, which is typically characterized as gaze-contingent blur \cite{guenter2012foveated,patney2016towards,tariq2022noise,tariq2024towards,krajancich2023towards,ashraf2025resolution,meng2018kernel,fan2024scene,stengel2016adaptive,zhang2024retinotopic,kaplanyan2019deepfovea, mohanto2025unified, groth2023wavelet}.
Gaze-contingent rendering has also been applied for display power reduction \cite{duinkharjav2022color,chen2024peapods}, for boosting stereo perception \cite{kergassner2025foveated, jimenez2026overdriving}, for reducing cybersickness~\cite{groth2024cybersickness, groth2021mitigation, groth2021visual}, or for improving neural rendering efficiency \cite{lin2025metasapiens,fovnerf}.
Another set of papers explores how asymmetric levels of blur in both eyes enhance foveated rendering \cite{wang2023deamp, meng2020eye,wang2025dominant}. While this asymmetry effectively creates a dichoptic percept, these works predominantly take advantage of eye dominance. Nevertheless, their results clearly show that stimuli with slight differences in blur can be fused. 
These findings motivate our work, but our approach is significantly different. Independent of eye dominance, we explore combinations of blur and sharpening operations on either eye that yield the highest level of modulation at the threshold of detectability, relying on binocular fusion and interocular blur suppression.



\section{Psychophysical Framework \feedbackNeeded}
\label{sec:experiments}
The core hypothesis of this experiment is that binocular fusion due to asymmetric blur--sharpen operations in each eye may preserve more visual detail than asymmetric dichoptic blur.
Similar to dichoptic tone-mapping techniques \cite{zhong2019dice}, if one eye retains higher-frequency structure (sharp) while the other eye provides a low-pass version of the same content (blur), the visual system may combine these asymmetric inputs into a fused percept that could appear closer to the reference.
Because visual acuity decreases with retinal eccentricity, we hypothesize that this blur--sharpen relationship \revise{is not uniform across the visual field}{is also eccentricity-dependent}.

\subsection{Method}
\label{subsec:blursharpen}
We parameterize both blur and sharpen operations by convolving images with Gaussian kernels. 
For a given input image $I$, the blurred image is produced as
\begin{equation}
I_{\mathrm{blur}} = G_{\blur} * I,
\end{equation}
where $G_{\blur}$ is a Gaussian kernel with standard deviation $\blur$, and $*$ is the convolution operator. 
In our implementation, we use a finite-support kernel whose size $\ksize$ is chosen to capture almost all of the Gaussian's mass,
\begin{equation}
\ksize = 2 \cdot \lceil 3\blur \rceil + 1.
\end{equation}
This definition is a common choice that truncates the filter at approximately $\pm 3\blur$\revise{~\cite{dali2026gaussian}}{}.
Sharpening is implemented via an unsharp masking-style operator. 
We first compute a blurred version of the original image using the same $\blur$, then compute the high-frequency residual $R$, and subsequently add this residual back to the original image with a multiplicative gain $\sharpen$,
\begin{equation}\label{Equ:sharpen}
\begin{aligned}
I_{\mathrm{sharp}} &=  I + \sharpen \cdot R \\
&= I + \sharpen \cdot (I - G_{\blur} * I).
\end{aligned}
\end{equation}
This formulation demonstrates how unsharp masking boosts the image components attenuated by the Gaussian low-pass filter. 
In practice, outputs are clamped to the valid display range.
\revise{}{However, clipping only occurs when the residual $R$ is large enough, which in practice happens for sharpening gains of $\phi > 1$. This constraint was accounted for when sampling parameters in our study.}

\begin{figure}
  \centering
  \includegraphics{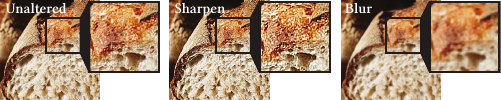}
  \Caption{
      Stimulus processing.
  }{
      We show a reference patch (left) after applying sharpening (middle) and blurring (right) of medium strength. Image by terski from Pixabay
  }
  \label{fig:stimuli_mainExp}
\end{figure}

We intentionally couple the sharpening scale with the blur scale by using the same Gaussian standard deviation, $\blur$, for both operations. 
If sharpening were the inverse operation of blurring, using the same $\blur$ could potentially offset the effect during binocular fusion.
In practice, unsharp masking is not an exact inverse, so this hypothetical cancellation is only approximate.
An example of these operations is shown in Figure~\ref{fig:stimuli_mainExp}.

\subsection{Experimental Setup}
\label{subsec:common-setup}
\begin{figure}
  \centering
  \includegraphics[width=\linewidth]{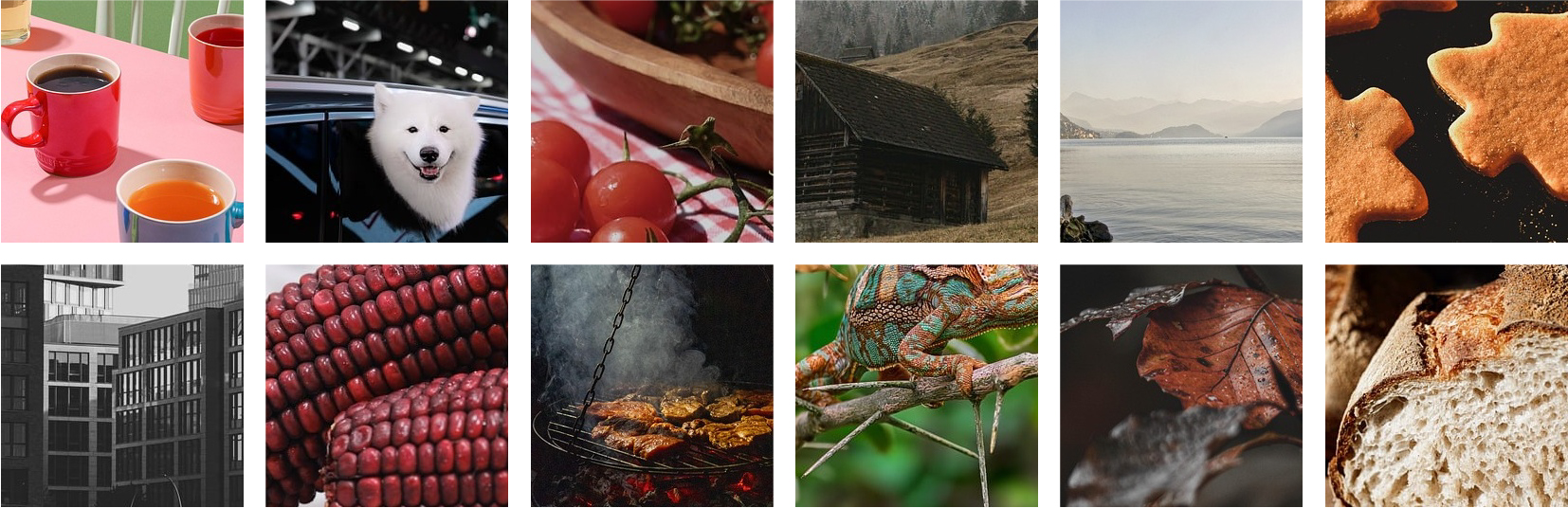}
  \Caption{
      Study stimuli.
  }{
      We showed twelve distinct image patches featuring a variety of image content, color profiles, and frequency of the content. Images from Pixabay.
  }
  \label{fig:patches}
\end{figure}

We used the same experimental setup for the pilot studies and the main perceptual study, which is first described here. 
\paragraph{Hardware apparatus}
Stimuli were presented on a Meta Quest Pro HMD.
The headset has a refresh rate of 90 Hz, 1,800$\times$1,920 resolution per eye, $106^\circ$ horizontal FOV, and $80^\circ$ binocular overlap.

\paragraph{Stimuli}
Our study was conducted on natural image patches sourced from Pixabay. The selected patches are shown in Figure~\ref{fig:patches}.
While perceptual studies often rely on controlled stimuli like Gabor patches, our goal was to study quality in a realistic setting.
%
We assigned each patch to a low, medium, or high spatial-frequency bin based on the relative amount of high-frequency energy in its Fourier power spectrum.
Each set of test stimuli contained two image patches from each frequency bin.
Patches were scaled to 141$\times$141 resolution for display, which was the smallest stimulus size that retains the highest spatial frequencies above the Nyquist limit of the headset. Patches were displayed on a solid gray background and had a size of approximately $7^\circ$ of visual angle. 
Patches were presented spanning equally-spaced steps across the headset's binocular field of view (eccentricities of $0^\circ$, $15^\circ$, or $30^\circ$) along the horizontal meridian, and the placement of stimuli on the left or right side was randomized across trials.
Observers were instructed to fixate on a persistent central crosshair; stimuli were hidden if gaze deviated.
The crosshair was removed when stimuli were presented at the foveal location ($0^\circ$).

\subsection{Pilot Experiments}
One critical caveat is that binocular fusion is not strictly guaranteed.
Sufficiently large interocular differences can lead to binocular rivalry or other forms of perceptual instability.
We therefore expect the fusion of dichoptic pairs to fail for some parameter ranges, depending on the blur--sharpen strength, the stimulus spatial frequency content, and retinal eccentricity.
For the main experiment, we conducted two pilot studies to identify a reasonable parameter range as well as find blur–sharpen levels with uniform perceptual spacing.

\subsubsection{Pilot 1: Calibrating Blur--Sharpen Range}
\label{subsec:pilot-1}
\begin{figure}
  \centering
  \includegraphics[width=\linewidth]{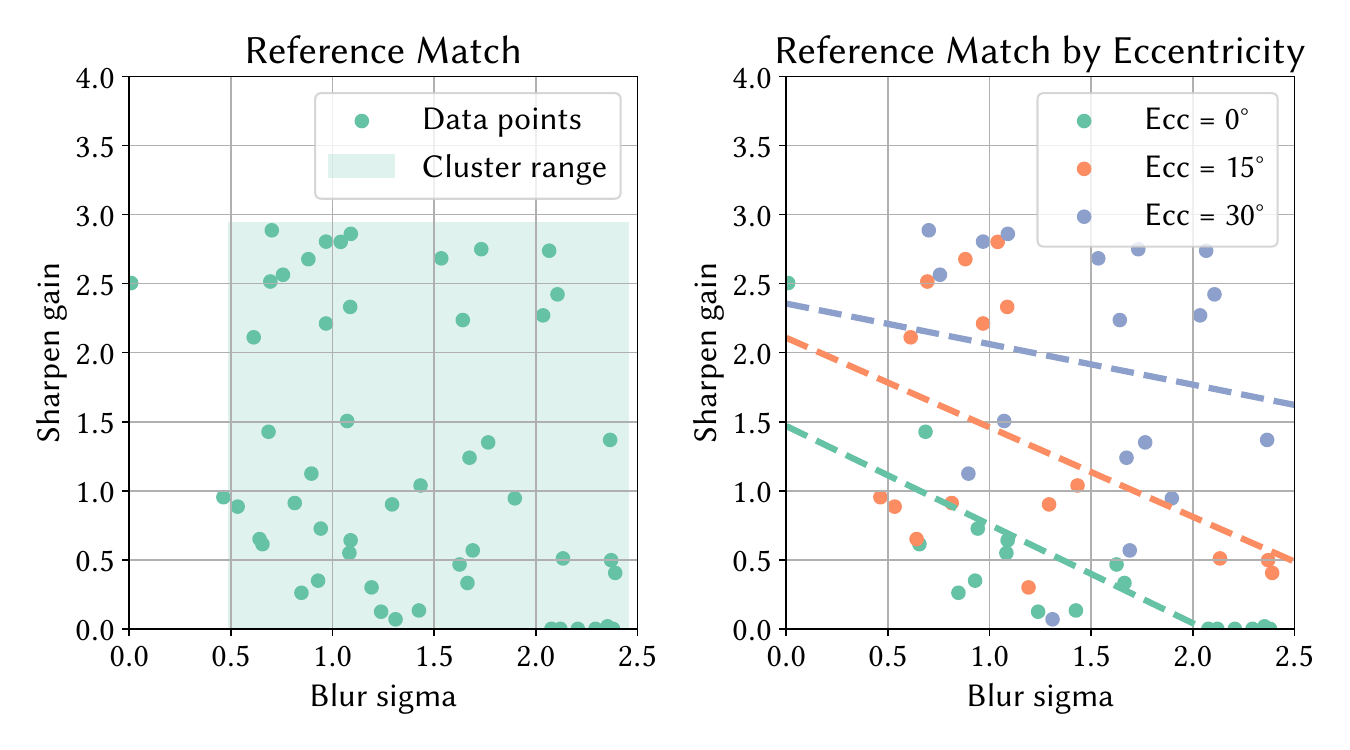}
  \Caption{
      Pilot study results.
  }{
      The left plot shows method of adjustment results of pilot 1. The results show that the selected parameters cluster well in the range outlined by the green background.
      The right plot shows the linear fits to per-eccentricity data, which were used to sample conditions for the 2nd pilot.
  }
  \label{fig:pilot_study_data}
\end{figure}

We conducted a method-of-adjustment~\cite{wichmann2001psychometric} study to identify a suitable parameter range for blur ($\blur$) and sharpening gain ($\sharpen$). 
We recruited 4 participants (1 female, 3 male, aged 22-31) for this pilot.
In each trial, they viewed a test dichoptic image pair placed at the three tested eccentricities, and were able to flip between this test stimulus and an unmodified reference image. 
The test dichoptic pair was initialized with a random combination of $\blur \in [0,3]$ and $\sharpen \in [0,4]$, and observers could adjust these two parameters until they found the fused percept to closely match the reference.
Across participants and images, the resulting adjustments clustered within a well-defined range of blur and sharpen values (see Figure~\ref{fig:pilot_study_data}). 
Based on these observations, we selected $\blur \in [0, 2.5]$ and $\sharpen \in [0, 2.5]$ as the range for subsequent testing.

\subsubsection{Pilot 2: Determining Main Study Parameters}
\label{subsec:pilot-2}
The first pilot established the parameter range, but left the question open on how best to sample this range to achieve perceptually-uniform spacing. While simple linear spacing would be the most straightforward parametrization, it might not result in a \textit{perceptually} uniform test set~\cite{wichmann2001psychometric}, leading to weak model approximation in the main study.

We ran a second pilot study to select specific stimulus levels that vary in approximately uniform perceived quality. 
Using the same stimulus pipeline, 5 observers (4 male, 1 female, aged 22--31 years), 
completed a two-alternative forced-choice (2AFC) similarity task with an unaltered reference.
Each trial presented two test dichoptic pairs (parameterized by $(\blur, \sharpen)$), where participants selected the test pair that appeared closer to the reference.
We evaluated a grid of parameter values by uniformly sampling five blur levels and five sharpening gains over the range determined in the first pilot $\blur, \sharpen\in[0, 2.5]$, forming a Latin grid for pairwise comparison.
The pairwise preference data were converted into a perceptual JOD scale via \textit{pwcmp}~\cite{perezortiz2017practicalguidesoftwareanalysing}.

To prioritize accurate sampling in foveal conditions, we used the $0^\circ$ condition and examined JOD scores as a function of $\blur$ and $\sharpen$ separately. 
We fit a linear model to these data to estimate the parameter value corresponding to 0 JOD, which corresponds to the reference condition (i.e., $\blur=0$ and $\sharpen=0$). 
We then selected four additional levels at 0.5 JOD decrements from this reference point, yielding five blur and five sharpening values that are approximately evenly spaced in perceived quality,
\begin{itemize}
    \item[] Blur ($\blur$): [0, 0.2704, 0.5408, 0.8113, 1.0817],
    \item[] Sharpen ($\sharpen$): [0, 0.2235, 0.4471, 0.6706, 0.8941].
\end{itemize}

\subsection{Main Experiment}
\label{subsec:main_study}

\begin{figure*}
  \centering
  \includegraphics[width=\linewidth]{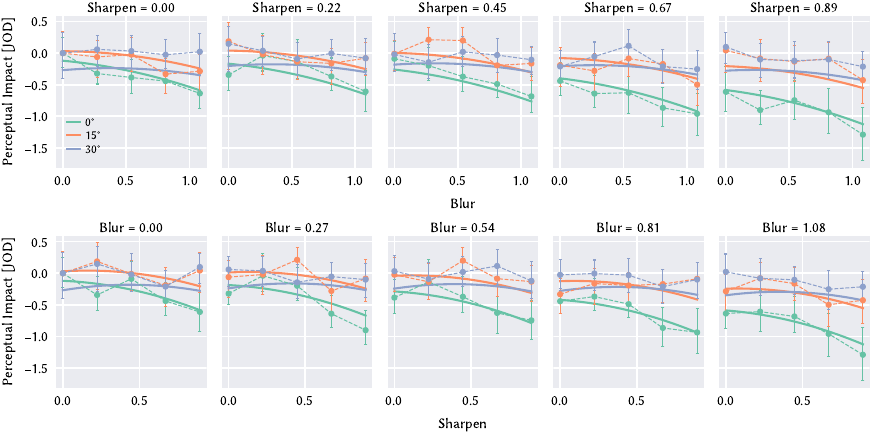}
  \Caption{
      Main study results.
  }{
      Plots visualizing the results of our perceptual study as a function of JOD. The plots show varying blur (top row), sharpening (bottom row), and eccentricity (color). The error on the datapoints represents the standard deviation, and the fitted quadratic functions are represented by solid lines.
  }
  \label{fig:main_study_data}
\end{figure*}

We conducted a within-subjects experiment to measure the perceived image quality of dichoptic pairs as a function of eccentricity and image content. The experiment used the same general experimental setup described in Section~\ref{subsec:common-setup}.

\paragraph{Task and procedure}
The experiment compared different dichoptic test stimuli against a binocular unaltered reference image in a 2AFC task.
Dichoptic test stimuli considered combinations of eccentricity, $\blur$, $\sharpen$, and image content. 
Each trial first presented the binocular reference and two dichoptic test stimuli.
Participants could freely toggle between the reference and two test stimuli, with a 500\,ms mid-gray blank screen inserted between views to prevent direct comparisons. Pairwise comparisons were scheduled using ASAP \cite{mikhailiuk2020active}, which samples stimuli by maximizing information gain.
This reduces the number of trials significantly compared to a full study design.
During each trial, the participant indicated which of the test stimuli appeared closer to the reference and made selections with a standard keyboard.
%
The experiment was divided into 15 min blocks with mandatory breaks between blocks due to the study duration (approximately one hour per participant).
Each participant completed 419 trials.

\paragraph{Participants}
A total of 31 participants (22 male, 9 female, aged 18-29 years) 
completed the experiment.
Participants reported normal or corrected-to-normal vision.
The study protocol was approved by the university’s institutional review board (IRB).
Participants gave informed consent before starting the experiment.

\subsection{Results}
We scaled the pairwise comparison responses to JODs using Thurstone's Case V model \cite{thurstone2017law}, as implemented by the \textit{pwcmp} algorithm \cite{perezortiz2017practicalguidesoftwareanalysing}.
Confidence intervals were estimated by simulating 500 bootstrap samples using the same \textit{pwcmp} algorithm.
The Kolmogorov-Smirnov test was run on the bootstrap distributions, where 64 of 75 conditions did not reject normality, suggesting that the normality assumptions are reasonable for most conditions.

We anchored JOD scores to the baseline condition (JOD=0 for $\blur=0$, $\sharpen=0$) so that they reflected perceptual impact at the given eccentricity.
We ran a three-way factorial analysis of variance (ANOVA) on anchored JOD scores across the main factors blur, sharpen, and eccentricity, including all two-way interactions.
The analysis showed strong main effects of blur ($p=5.44\cdot10^{-8}$, $F(4,32) = 18.61$), sharpen ($p=6.32\cdot10^{-8}$, $F(4,32) = 18.35$), and eccentricity ($p=5.77\cdot10^{-16}$, $F(2,32) = 127.39$). The interaction effect between blur and sharpen was not significant ($p=0.707$, $F(16,32) = 0.77$). However, the interaction between blur and eccentricity ($p=0.03$, $F(8,32) = 2.45$), and between sharpen and eccentricity was significant ($p=5.38\cdot10^{-5}$, $F(8,32) = 6.45$), indicating that eccentricity modulates the impact of the blur--sharpen fusion (see Figure~\ref{fig:main_study_data}). 
\revise{}{These interaction effects are consistent with the known fall-off of visual acuity with eccentricity, creating non-uniform dichoptic effects across the visual field.}


\section{Computational Model\feedbackNeeded}
\label{sec:modeling}
\begin{figure*}
  \centering
  \includegraphics{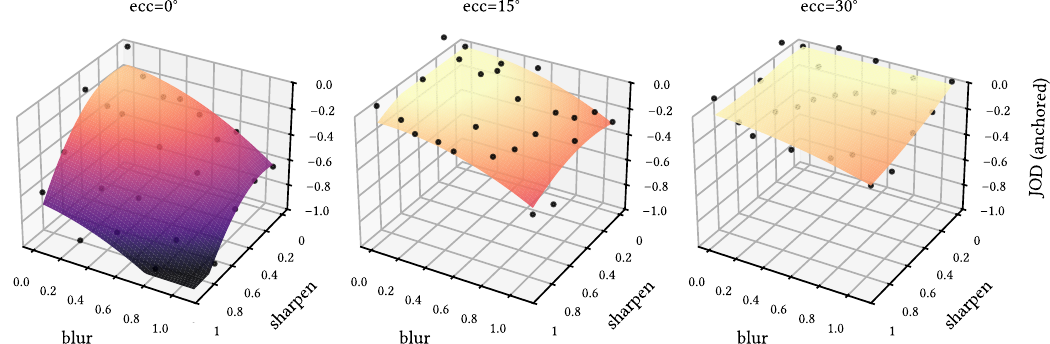}
  \Caption{
      Perceptual model.
  }{
      Visualization of the quadratic model surface for each eccentricity, respectively. The results show how sensitivity to dichoptic effects decreases with eccentricity.
  }
  \label{fig:model_fit}
\end{figure*}

To interpolate our data, we fit 4D perceptual models that estimate perceptual impact (JOD scores) given blur ($\blur$), sharpen ($\sharpen$), and eccentricity ($\ecc$) as input (see Figure~\ref{fig:model_fit}).
We considered four parameterizations: linear, linear with pairwise interactions, exponential, and quadratic.
Among these, the quadratic model $\fquadratic(\cdot)$ yielded the best numerical fit ($R^2 = 0.82$). It is expressed as
\begin{equation}
\begin{aligned}
\fquadratic(\blur,\sharpen,\ecc) =\, &\param_1 + \param_2\blur + \param_3 \sharpen + \param_4 \ecc
+ \param_5 \blur^2 \\ + &\param_6 \blur\sharpen + \param_7 \blur\ecc + \param_8 \sharpen^2 + \param_9 \sharpen\ecc + \param_{10} \ecc^2.
\label{eq:quadratic_model}
\end{aligned}
\end{equation}
The parameters, $\param_i \in \mathbf{\param}$, were fitted to the empirical results of the perceptual experiment (see Section~\ref{subsec:main_study}). The optimized parameters are $\mathbf{\param} = [-.12,-.189, -.109,.025,-.226,-.074,.012,-.459,.017,-.001]$.

Practical application of a perceptual model also relies on the ability to extrapolate beyond the tested regime.
In our analysis, the linear model ($R^2 = 0.65$), $\flinear(\cdot)$,
\begin{equation}
\begin{aligned}
\flinear(\blur,\sharpen,\ecc) &= \param_1 + \param_2 \blur + \param_3 \sharpen + \param_4 \ecc,
\label{eq:linear_model}
\end{aligned}
\end{equation}
extrapolated most accurately to values outside the experimental parameter space. This was most pronounced when exploring higher eccentricities (cf. Section \ref{sec:application}).
The fitted parameters were \[\mathbf{\param} = [-.179, -.284, -.301,.016].\]

\section{Application to Foveated Rendering \feedbackNeeded}
\label{sec:application}

While our model offers fundamental insights into the neurological processing of stereoscopic information, it also enables practical applications in computer graphics. 
A direct application of our model is the optimization of foveated rendering.
By rendering the foveated image using dichoptic blur--sharpen image processing, we argue that visual quality can be better preserved even when fewer samples are used.
These two objectives of equal perceptual quality and lower sampling requirements are validated separately.

%


\subsection{Implementation \& Sampling}\label{sec:sampling}

To validate the quality objective, we implemented dichoptic foveated rendering through post-processing shaders in Unity. 
By applying blurring and sharpening in post-processing, we directly simulated the visual output of an undersampled-then-reconstructed frame, providing a rigorous testbed for evaluating our model’s ability to preserve fidelity in high-resolution, dynamic environments.

To determine the optimal blur ($\blur$) and sharpening ($\sharpen$) parameters for a given pixel at eccentricity $\ecc$, we used the linear model $\flinear(\cdot)$, as it was found to be more robust with respect to extrapolation (see Section~\ref{sec:modeling}).
This property is essential for covering the full range of eccentricities in wide-FOV HMDs.
Because the linear model yields a planar objective surface, the optimization space lacks local minima, precluding the need for iterative optimization solvers; instead, we adopted an analytical approach to sampling.

Our primary objective was to maximize the blur magnitude $\blur$, as it directly influences the reduction in required samples and thus yields computational savings during rendering (see Section~\ref{subsec:compute_benefit}). In contrast, the computational overhead of sharpening is constant.
As shown in Equation~\ref{Equ:sharpen}, sharpening is implemented via a single addition and multiplication. The $O(1)$ complexity renders its cost negligible relative to rendering costs (see Section~\ref{subsec:blursharpen}), ensuring that the computational benefits gained from increased blur are not offset by additional pipeline complexity.

Our model revealed an inverse relationship between sharpening strength and blur magnitude.
Since human sensitivity to high-frequency sharpening artifacts diminishes significantly at higher eccentricities, we hypothesized that sharpening gain could be decreased with eccentricity to permit greater peripheral blur.
However, in the fovea, visual quality must be better preserved due to higher acuity and to prevent binocular rivalry.
We therefore assigned the highest sharpening gain validated in our experiments ($\sharpen = 0.89$) to the fovea.
This high sharpening minimized the required blur at the gaze center while still providing meaningful computational savings compared to standard rendering.

In our experiment, we tested two conditions: a sharpening gain that is constant across the visual field, and one that decreases linearly to the lowest validated sharpen value ($\sharpen = 0.22$) between $0^\circ$ to $60^\circ$ eccentricity.
The corresponding blur values are derived from sampling the model at $-1$ JOD.

\revise{}{Recent stereo-aware rendering work has shown that exploiting inter-view coherence can improve stereo consistency while keeping computational overhead low~\cite{philippi2025stable}.}
Our approach is designed to complement, rather than replace, existing foveated rendering frameworks.
In this application, dichoptic sharpening operates exclusively on the low-pass information that remains after classic foveation has been applied. High-frequency information beyond the Nyquist limit of foveated rendering is not visible to the viewer and may even introduce aliasing. By only sharpening on the foveated image, the computational cost of dichoptic rendering never exceeds that of foveated rendering. 


\subsection{Baseline Calibration}

\begin{figure*}
  \centering
  \includegraphics{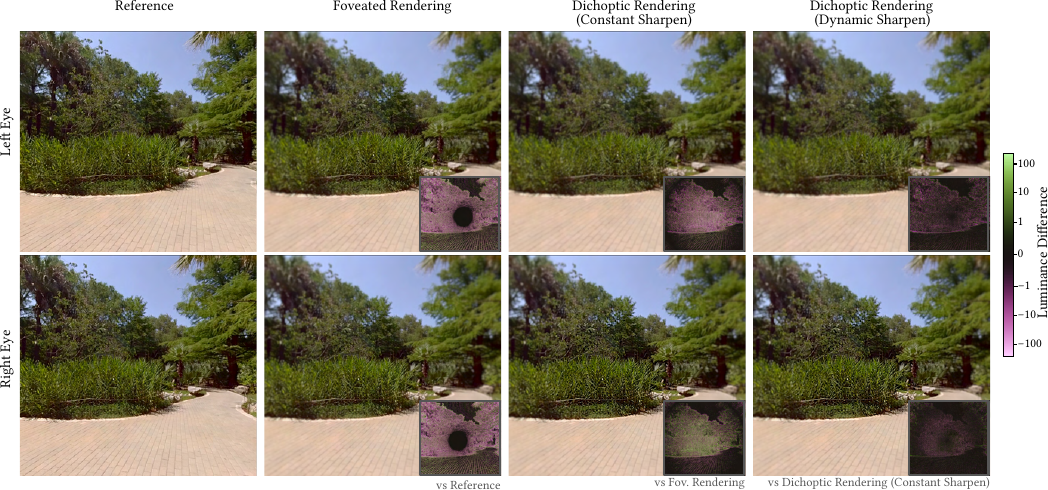}
  \Caption{
      Application renderings.
  }{
      Visual stimuli of reference, foveated rendering, and the two dichoptic rendering approaches as displayed in the validation study. The insets highlight luminance differences between methods for better visibility. \copyright The University of Texas at Austin.
     
  }
  \label{fig:validation_stimuli}
\end{figure*}

Prior foveated rendering methods are usually calibrated by a central, fixed gaze point~\cite{patney2016towards, tariq2022noise,tariq2024towards, kergassner2025foveated, walton2021beyond}. 
In contrast, we apply a dynamic gaze setting, which requires recalibration to account for artifacts due to saccadic eye movements.
We calibrated the baseline foveated rendering parameters following the formulation of \citet{tariq2022noise},
\begin{align}
\blur = c \cdot \max(\ecc - 8, 0).
\label{eq:fov_rendering}
\end{align}
To calibrate the foveation intensity, $c$, we conducted a 2AFC study (N=14, 9 male, 5 female, aged 20-31)
with three 360° stereoscopic videos from the LIVE Database~\cite{jin2020live}. We chose diverse videos with natural content, people, and a street with cars. Each trial compared stimuli with foveated rendering applied with one of four intensities ($c \in [0, 1.5]$) to their unmodified reference.
Each comparison was repeated three times per video, resulting in 36 trials per participant.

We analyzed participants' accuracy (\% of reference chosen) as a function of foveation intensity $c$ and fit a logistic psychometric function to the data.
Performance increased reliably with blur; the psychometric fit performed better than a constant model (bootstrap $p = 3.33 \cdot 10^{-4}$) and showed no evidence of poor fit (bootstrap deviance $p = 0.197$).
We obtained the perceptual threshold (-1 JOD) from the fitted curve, which was $c = 0.11$ with a parametric-bootstrap 95\% confidence interval of (0.09, 0.14).
The data and fitted curve are visualized in the supplementary webpage.

\subsection{Subjective Quality Evaluation}
\label{subsec:validation_evaluation}
We conducted a within-subjects repeated-measures experiment to rate the perceptual quality of the baseline reference, the calibrated foveated rendering, and two conditions for dichoptic foveation on high-resolution stereoscopic 360° videos in VR (see Figure~\ref{fig:validation_stimuli}). 
The two dichoptic conditions applied constant and dynamic sharpening, respectively, as described in Section~\ref{sec:sampling}.
Immersive video content allowed us to test the model’s ability to generalize to full-screen, dynamic scenes.

\paragraph{Apparatus and stimuli}
The Meta Quest Pro HMD was used.
Stimuli consisted of ten $360^\circ$ stereoscopic videos at a resolution of $7680 \times 3840$, obtained from the LIVE Database~\cite{jin2020live}.
Each participant viewed all videos under all rendering modalities in randomized order.

\paragraph{Participants and procedure}
A total of 14 participants (8 male, 5 female, 1 other, aged 18--26 years)
completed the experiment. 
Similar to the main study, participants were screened for normal vision and compensated for their time.
Each trial presented a ten-second 360° video in which participants were allowed to freely explore the scene.
After each video finished playing, participants rated video content using a 5-point Likert scale which conformed to the ITU-T P.910 standard for video quality assessment~\cite{itu2023subjective}.

\subsection{Results}
\begin{figure}
  \centering
  \includegraphics{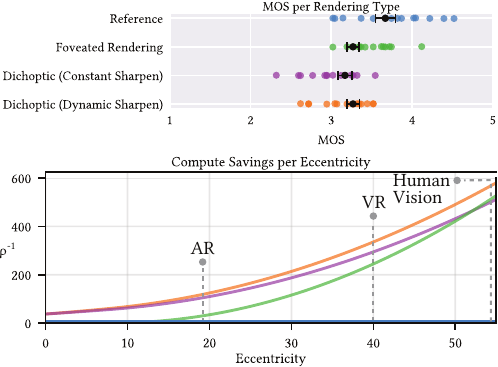}
  \Caption{
      Validation result.
  }{
      The top graph visualizes perceptual quality for the investigated techniques in MOS. Higher values indicate better quality. The dots depict subject-level ratings, while the weighted mean with standard deviation is shown in black.
      The bottom graph shows computational benefits measured by the inverse of sampling density (gain factor $\rho^{-1}$, $y$-axis) as a function of eccentricity ($x$-axis). Colors correlate with the methods of the top graph. The binocular FOV limits of common devices are also shown. 
  }
  \label{fig:validation_results}
\end{figure}

The user ratings were scaled to mean opinion scores (MOS) using the Sureal method~\cite{netflix2016sureal}.
The results can be seen in Figure~\ref{fig:validation_results}.
The mean MOS values for dichoptic rendering with constant sharpening (3.16), dynamic sharpening (3.26), and standard foveated rendering (3.26) were similar, whereas the reference condition received a slightly higher score (3.66).
To analyze significance of these results, we first tested the subject-level MOS for normality using Shapiro-Wilk ($W=0.9856$, $p=0.738$), which affirmed this assumption.
A repeated-measures ANOVA was then conducted to determine if the difference in MOS between the render types was significant ($p<0.0001$, $F=9.59$). Given the confirmation of the significance hypothesis, we followed with post-hoc pairwise comparisons using t-tests with Bonferroni correction.
%
We found that only dichoptic foveated rendering with constant sharpening was significantly different from the reference ($p=0.015$). Both dynamic sharpening and foveated rendering showed no significant difference with the reference video in terms of visual quality. This result suggests that our model generalizes well beyond simple image patch stimuli to dynamic, full-screen content.
%

\subsection{Computational Benefit}
\label{subsec:compute_benefit}
We demonstrated that dichoptic foveation achieves similar visual quality compared to full-resolution rendering, despite significant application of blur.
The computational benefit of our approach is measured using a theoretical, hardware-agnostic framework based on blur kernel strength, $\blur$. 
We consider foveated raytracing, where the primary objective is to maximize the local sample spacing $\Delta x(\ecc)$ relative to retinal eccentricity $\ecc$. 
In this framework, any $\Delta x(\ecc) \ge 1$ represents a reduction in local sampling density compared to full-resolution rendering; therefore, larger values of $\Delta x$ directly correspond to greater computational savings.

First, we define the relationship between blur $\blur$ and the spatial sampling interval $\Delta x$. 
In alignment with standard practices in the rendering literature, we assume that sparse sampling followed by an ideal reconstruction filter behaves as a low-pass operation \cite{albert2017latency,hoffman2018limits}. Specifically, we approximate the resulting point-spread function (PSF) as a Gaussian kernel.
The frequency domain expression of the PSF \cite{goodman2005introduction} with standard deviation $\blur$ is defined as
\begin{equation}
    H(f)=e^{-2\pi^2\blur^2f^2}.
\end{equation}
In rendering, the practical limit of frequency $f$ is given by the Nyquist theorem where the spatial frequency drops below 0.5 cycles per pixel. Since a Gaussian has no true cutoff, we set $f_c = \frac{1}{2\pi\blur}$ as the natural scale frequency, where the Gaussian’s energy is concentrated, given its spatial width, $\blur$.
To reproduce this blur without additional loss, the sampling rate must satisfy
\begin{equation}
f_s \ge 2f_c = \frac{1}{\pi\blur}.
\end{equation}
Given that a frequency in pixel space is naturally related to the sampling spacing by $f_s = \frac{1}{\Delta x}$, we then have
\begin{equation}
\Delta x = \pi\blur.
\end{equation}
For 2D images, the sample density relative to full resolution is
\begin{equation}
\rho = \frac{1}{\Delta x^2} = \frac{1}{(\pi\blur)^2}.
\end{equation}
Figure~\ref{fig:validation_results} shows the theoretical computational savings that can be achieved by foveated and dichoptic rendering relative to uniform sampling as a function of eccentricity. 

Computational savings for the full display are computed by aggregating local eccentricity-dependent savings across the entire FOV. We assume central gaze and approximate the display with a circular area. With these assumptions, savings are integrated across the spherical domain. Because the solid angle of each concentric ring increases with eccentricity, the base saving function must be weighted by the area of the corresponding spherical cap. We express the total theoretical computational cost, $C$, using the angular sampling formulation,
\begin{equation}
C = 2\pi \int_{0}^{\ecc_{\max}} f(\ecc) \sin(\ecc) \, d\ecc.
\end{equation}
In this expression, $f(\ecc)$ represents the local sampling density at eccentricity $\ecc$. 
We evaluate the cost for both standard foveated rendering ($C_\text{fr}$) and our proposed dichoptic foveation ($C_\text{dr}$) to derive a ratio of their respective workloads, $\frac{C_\text{dr}}{C_\text{fr}}$. 
For the VR HMD that was used in our study, the analysis yields a benefit of approximately 1.9, suggesting that dichoptic foveation requires only about half the samples of traditional foveated rendering to maintain the same perceived quality.
For smaller stereoscopic displays, such as those used in augmented reality (AR), dichoptic foveation yielded theoretical savings of up to 11$\times$.
Additional device-specific results are provided in the supplementary material.

\subsection{Ablation: Dichoptic Rendering Without Eyetracking}
\begin{figure}
  \centering
  \includegraphics{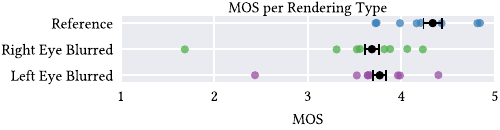}
  \Caption{
      Ablation result.
  }{
    The jitter plot compares the perceived quality (MOS) of reference videos to uniform dichoptic rendering. For the analysis, we split the results into stimuli that blurred the right and left eye respectively. Greater MOS indicates better quality. 
  }
  \label{fig:ablation_results}
\end{figure}

Many modern devices lack eye tracking.
As a simple adaptation, we evaluated the application of dichoptic parameters derived at zero eccentricity uniformly across the image, eliminating the need for gaze-contingent rendering.
Our results showed that even in the contrast-sensitive fovea, substantial dichoptic modulation was tolerated.
Under uniform dichoptic rendering, theoretical sampling requirements were reduced by approximately $97\%$ relative to native-resolution rendering.

To validate the perceptual quality of this approach, we conducted a small-scale ablation study (9 participants, 6 male, 3 female, aged 19--26 years),
where participants rated the video quality of the uniform dichoptic rendering method against the unaltered reference.
Similar to Section~\ref{subsec:validation_evaluation}, users conducted a Likert-based rating task across ten $360^\circ$ videos.
Results are shown in Figure~\ref{fig:ablation_results}.
To explore potential ocular dominance effects, we analyze our results in two conditions: blurring the left and right eye respectively.
A repeated-measures ANOVA indicated a significant effect of condition ($p=0.003$).
Post-hoc comparisons showed that when the left eye was blurred, the dichoptic condition ($MOS=3.63$) was not significantly different from the reference ($MOS=4.19$) after Bonferroni correction ($p=0.04$). These results indicate that uniform dichoptic rendering can indeed achieve a perceptual quality comparable to native rendering.
In contrast, when the right eye was blurred, results indicated a significant difference from the reference ($p=0.06$).
This asymmetry may be related to eye dominance, as
nearly 70\% of people have been found to be right-eye dominant~\cite{seijas2005eye}. Preserving sharpness in the dominant eye may therefore yield higher perceived quality, highlighting an important direction for future work.

\section{Limitations \& Future Work \feedbackNeeded}
\label{sec:limitations}

The objective of this paper is to define a general approach to dichoptic foveation that applies to a typical observer.
As such, we computed average JOD scores across all users in our psychophysical study; individual differences in sensitivity could lead to variation in perceived quality or the potential for binocular rivalry.
A per-observer calibration procedure could therefore improve subject-level performance and represents a promising direction for future work.

Our model does not explicitly account for individual eye dominance or dynamic effects like change blindness~\cite{groth2023instant}, which may offer additional opportunities for optimization. In this study, the assignment of blur and sharpening filters was randomized across the left and right eyes to ensure the resulting computational model remains broadly applicable. However, as suggested by our ablation study, the effectiveness of dichoptic foveation may depend on which eye receives the sharper signal.
Future work could incorporate eye-dominance-aware rendering strategies to enable more personalized dichoptic foveation.

All experiments were conducted using a commercial HMD (Meta Quest Pro), which introduces limitations related to eye-tracking latency, display resolution, optical distortions, and related factors.
Such limitations may introduce auxiliary artifacts that are not directly related to the phenomena studied in this work.
Nevertheless, our model is calibrated under conditions that are representative of a realistic VR viewing scenario.

Dichoptic effects, furthermore, require binocular vision. However, typical VR displays have a monocular region in the outer edges of the visual field, where there is no overlap between the two eyes.
Our technique, however, only works in this overlapping region; headsets with very high FOV may have a smaller overlapping region relative to the overall display area, leading to less savings.
In this high-FOV monocular regime, normal foveated rendering can be applied to supplement dichoptic foveation.

Our application-focused evaluation assessed perceptual quality relative to a baseline reference condition. Therefore, long-term effects like discomfort due to eye strain or rivalry are unexplored, and should be investigated as future work when used in daily use cases.

The aim of our computational analysis was to be system-agnostic. However, real-world raytracing systems come with system-level optimizations that may lead to deviations when compared with our theoretical savings. Future work should therefore implement our approach on such a system to evaluate actual hardware-specific savings. \revise{}{Other techniques, such as those targeting bandwidth-reduction, could supplement our method for real-time stereo transmission~\cite{denes2019temporal}}.

Additional future directions include exploring a wider range of eccentricities or blur--sharpen parameters, as well as additional interaction effects due to contrast or color.
Furthermore, alternative display modalities like AR may realize even more gains with our technique, due to low FOV (see Figure~\ref{fig:validation_results}) and reduced contrasts that could mask distortions.


\section{Conclusion}
\label{sec:conclusion}

This work introduces \textit{dichoptic foveation}, a paradigm that leverages binocular fusion to optimize stereoscopic rendering. Psychophysical measurements show that appropriate combinations of blur and sharpening, parameterized by retinal eccentricity, preserve or improve perceived quality relative to symmetric presentations. The resulting 4D perceptual model provides a compact and interpretable quality predictor that directly links rendering parameters to human perception. Applied to foveated rendering in VR, the model enables \textasciitilde50\% sampling-rate reductions compared to traditional foveated rendering, while maintaining perceptually equivalent quality to full-resolution renderings.

\begin{acks}
    This work is partially supported by National Science Foundation grant \#2225861, and the DARPA Intrinsic Cognitive Security (ICS) program.
\end{acks}

\bibliographystyle{templates/acmart/ACM-Reference-Format}
\bibliography{bibliography}

\newpage 



\end{document}